\documentclass[a4paper, amsfonts, amssymb, amsmath, apssamp, nofootinbib, reprint, prl, aps, showkeys, twoside,superscriptaddress,noeprint]{revtex4-2}

\usepackage[english]{babel}

\usepackage{physics}

\usepackage{graphicx}
\usepackage{dcolumn}
\usepackage{bm}
\usepackage{xcolor}
\usepackage{ulem}
\usepackage{comment}

\begin{document}

\preprint{APS/123-QED}

\title{Superbandwidth laser pulses in a dispersive medium: oscillating beyond the Fourier spectrum with unexpected propagation features.}



\author{Enrique G. Neyra}
\email{enriqueneyra@cnea.gob.ar}
   \affiliation{Instituto Balseiro (Universidad Nacional de Cuyo and Comisión Nacional de Energía Atómica) and CONICET CCT Patagonia Norte. Av. Bustillo 9500, Bariloche 8400 (RN), Argentina. }

\author{Laureano A. Bulus Rossini}
   \affiliation{Instituto Balseiro (Universidad Nacional de Cuyo and Comisión Nacional de Energía Atómica) and CONICET CCT Patagonia Norte. Av. Bustillo 9500, Bariloche 8400 (RN), Argentina. }

\author{Fabi\'an Videla}
\affiliation{Centro de Investigaciones \'Opticas (CICBA-CONICET-UNLP), Cno.~Parque Centenario y 506, P.O. Box 3, 1897 Gonnet, Argentina}
\affiliation{Departamento de Ciencias B\'asicas, Facultad de Ingenier\'ia UNLP, 1 y 47 La Plata,Argentina}
             
\author{Lorena Reb\'on}
\email{rebon@fisica.unlp.edu.ar}
\affiliation{ 
Departamento de F\'isica, FCE, Universidad Nacional de La Plata, C.C. 67, 1900 La Plata, Argentina
}%
\affiliation{Instituto de F\'isica de La Plata, UNLP - CONICET, Argentina}

\author{Pablo A. Costanzo Caso}
   \affiliation{Instituto Balseiro (Universidad Nacional de Cuyo and Comisión Nacional de Energía Atómica) and CONICET CCT Patagonia Norte. Av. Bustillo 9500, Bariloche 8400 (RN), Argentina. }

\date{\today}

\begin{abstract}


The concept of superbandwidth refers to the fact that a band-limited signal can exhibit, locally, an increase of its bandwidth, i.e., an effective bandwidth greater than that predicted by its Fourier transform. In this work, we study
the propagation of superbandwidth laser pulses in a dispersive medium, characterized by the group velocity dispersion. In particular, two important results arise from the analysis of the instantaneous frequency of the pulse obtained through the Wigner function distribution: First, it can be observed local oscillations of the electric field which are beyond the Fourier spectrum of the incoming pulse. Second, for a range of values of the pulse synthesis parameters, surprisingly, the dynamics of the instantaneous frequency within certain temporal regions, corresponds to that of a pulse propagating in a medium with a group velocity dispersion of opposite sign. This phenomenon is intrinsic to the special characteristics of the pulse and not to the dispersive properties of the medium. 
\end{abstract}

\maketitle




{\textit{Introduction}}.---Wave phenomena are perhaps the most frequent physical phenomena in nature.
They are present in all branches of physics, from quantum mechanics to general relativity and, together with the superposition principle, plays a key role in the description of the universe. 
So it is that, in recent years, the wave phenomenon known as superoscillations has gained attention in the area of classical optics and beyond~\cite{berry2019roadmap}.

Superoscillations occur when a band-limited function locally oscillates faster than the highest frequency in its Fourier spectrum, which can be accomplished by properly manipulating the spectral phase of such a function to cause destructive interference between Fourier components.
This phenomenon was initially studied in a purely mathematical framework~\cite{berry2006evolution,aharonov2017mathematics}, to later find real applications to subdiffractive beams~\cite{zheludev2022optical,cheng2022super,wu2019broadband,zhang2023generating}, signal processing~\cite{ferreira2006superoscillations,luo2023creation}, ultrashort pulses~\cite{eliezer2017breaking,mccaul2023superoscillations}, acoustic waves~\cite{Brehm2020,shen2019ultrasonic}, among others~\cite{chen2019superoscillation,lin2023reconfigurable,dennis2008superoscillation,yuan2016quantum,eliezer2014super}.

In this context, in Ref.~\cite{neyra2021tailoring}, we have presented two equivalent techniques to obtain sub-diffractive Gaussian beams through the destructive interference of two pulses with different spatial widths. After considering the mathematical equivalence between Gaussian beams and ultra-short pulses, the same idea was implemented to obtain sub-Fourier ultra-short pulses~\cite{neyra2022simple}. In the temporal domain, these techniques generate an ultra-short pulse with a central temporal lobe whose full width at half maximum (FWHM) is below the Fourier limit. Subsequently, in Ref.~\cite{Neyra:21}, we introduced the wave phenomenon of {\it{superbandwidth}} (SB), in which these sub-Fourier laser pulses interact locally with matter as if they had a spectral bandwidth greater than that predicted by the Fourier transform. In other words, the SB pulse interacts as if it had frequencies higher and lower than those present in its Fourier spectrum~\cite{Neyra22}. 



As is known, when a laser pulse $E(t)$ propagates in a dispersive media characterized by the real part of the frequency-dependent refractive index $n(\omega)$, it spreads temporarily since each frequency ``within" its spectrum travels to a different speed. At first order, this phenomenon is described by the group velocity dispersion (GVD)~\cite{diels2006ultrashort}. In the case where the medium has a positive dispersion, i. e. $\mathrm{GVD}>0$, 
lower frequencies travel faster than the higher frequencies. Oppositely, if $\mathrm{GVD}<0$, higher frequencies travel faster than the lower ones. 

The possibility of controlling and manipulating both the sign and value of GVD through different devices such as prisms~\cite{fork1984negative}, diffraction gratings~\cite{strickland1985compression} or chirped mirrors~\cite{kartner1997design}, is essential in ultrafast optics~\cite{weiner2011ultrafast,walmsley2001role} and strong field laser physics~\cite{brabec2008strong}. 
Also, a dispersive medium allows to decompose, over the time, the spectral content of a Fourier-limited laser pulse when it travels in the medium. Moreover, taking into account the first relevant approximation for a dispersive media i.e. the GVD, when the pulse propagates a long distance, the GVD gives the temporal Fourier transform~\cite{Jannson1983,Goda2013}.



In this work, we theoretically studied the propagation of a SB laser pulse in a dispersive medium. Its effect on a propagating pulse is represented by a linear operator which introduces a quadratic phase 
\begin{equation}
H_D(\omega)=e^{-i\Omega(\omega-\omega_0)^2},
\end{equation}
where $\Omega=\frac{1}{2}\mathrm{GVD}\times z$ and $z$ is the propagation distance. Although the medium is linear 
and the synthesis of the SB pulse 
can be obtained by a simple experimental technique, 
the instantaneous frequency of 
such a pulse, $\omega(t)=\frac{d\phi(t)}{dt}$, as it travels through  
the medium, exhibits a very complex and physically rich behavior.
Firstly, we will report the existence of local oscillations of the electric field, represented by $\omega(t)$, which appear when the SB pulse propagates in the medium.  
Within some temporal window, 
these oscillations are beyond the Fourier spectrum of the original pulse. Secondly, we will show that, in a range of values 
of the parameters that controls the synthesis of the SB pulse, surprisingly, 
$\omega(t)$ behaves as if the pulse were traveling in a medium with a GVD of opposite sing.  
Both effects disappear when the propagation distance is large enough, 
and the frequency $\omega(t)$, as given by the Fourier spectrum, is recovered.


{\textit{Results}}.---We will start by defining the SB pulses analyzed in this work, which are similar to those studied in Ref.~\cite{Neyra22}. These pulses arises from the destructive interference between two Gaussian pulses with different amplitudes and temporal widths, which can be easily synthesized using an interferometric system. In the spectral domain, the SB pulses are described by the expression
\begin{equation}
\tilde{E}_{SB}(\omega)=e^{-(\frac{\omega-\omega_0}{\Delta\omega})^2}-\alpha e^{-(\frac{\omega-\omega_0}{\beta\Delta\omega})^2},\label{e1}
\end{equation}
%
where the first term $\tilde{E}_{G}(\omega)=e^{-(\frac{\omega-\omega_0}{\Delta\omega})^2}$ corresponds to the electric field of the initial Gaussian pulse, normalized to the unit, which travels through the interferometer without being modified.
In this context, $\omega_0$ and $\Delta\omega$ denote its carrier frequency and bandwidth, respectively. The second term in Eq.~\eqref{e1}, stands for the pulse that was modified before it is recombined. Here, $\alpha$ indicates the amplitude ratio between these two interfering pulses, while $\beta$ (0$<\beta<$1) is the relative temporal width that can be controlled, experimentally, by means of a spectral Gaussian filter in one of the interferometer paths.
A theoretical analysis on the reduction of the temporal FWHM of the central lobe of the SB pulse, as a function of $\alpha$ and $\beta$, can be found in Ref.~\cite{neyra2021tailoring}. 
Consequently, the synthesized SB pulse when propagating through a dispersive medium, can be expressed as: 
\begin{eqnarray}
 E_{SB}^D(t)\equiv |E_{SB}^D(t)| e^{i\phi(t)} &=& \mathcal{F}[\tilde{E}_{SB}^D(\omega)]\\ &=&\mathcal{F}[\tilde{E}_{SB}(\omega)e^{-i \Omega(\omega-\omega_0)^2}].\nonumber
\label{e2}
\end{eqnarray}
%
%

%

In what follows, we will resort to the Wigner function formalism that can be apply to classical optics to represent the state of a light field~\cite{alonso2011wigner}, and in particular, it has been used for a time-frequency analysis to study the dynamics of pulse propagation through dispersive media~\cite{Azana2005,Ojeda2007}.
For instance, starting from the representation of the field in the frequency domain, 
the Wigner function distribution in the time-frequency phase space is defined as
\begin{equation}
 W(t,\omega)=\frac{1}{2\pi}\int_{-\infty}^{\infty} \tilde{E}(\omega+s/2)\tilde{E}^*(\omega-s/2)e^{i t s} ds,
\label{e5}
\end{equation}
from where, by replacing the mathematical expression of the field under consideration, the instantaneous frequency of such a pulse can be obtained as the average:
%
\begin{equation}
 \omega(t)=\frac{\int_{-\infty}^{\infty}\omega W(t,\omega)d\omega}{\int_{-\infty}^{\infty} W(t,\omega)d\omega}~.  
\label{e6}
\end{equation}

\begin{figure}[h!]
\includegraphics[width=0.4\textwidth]{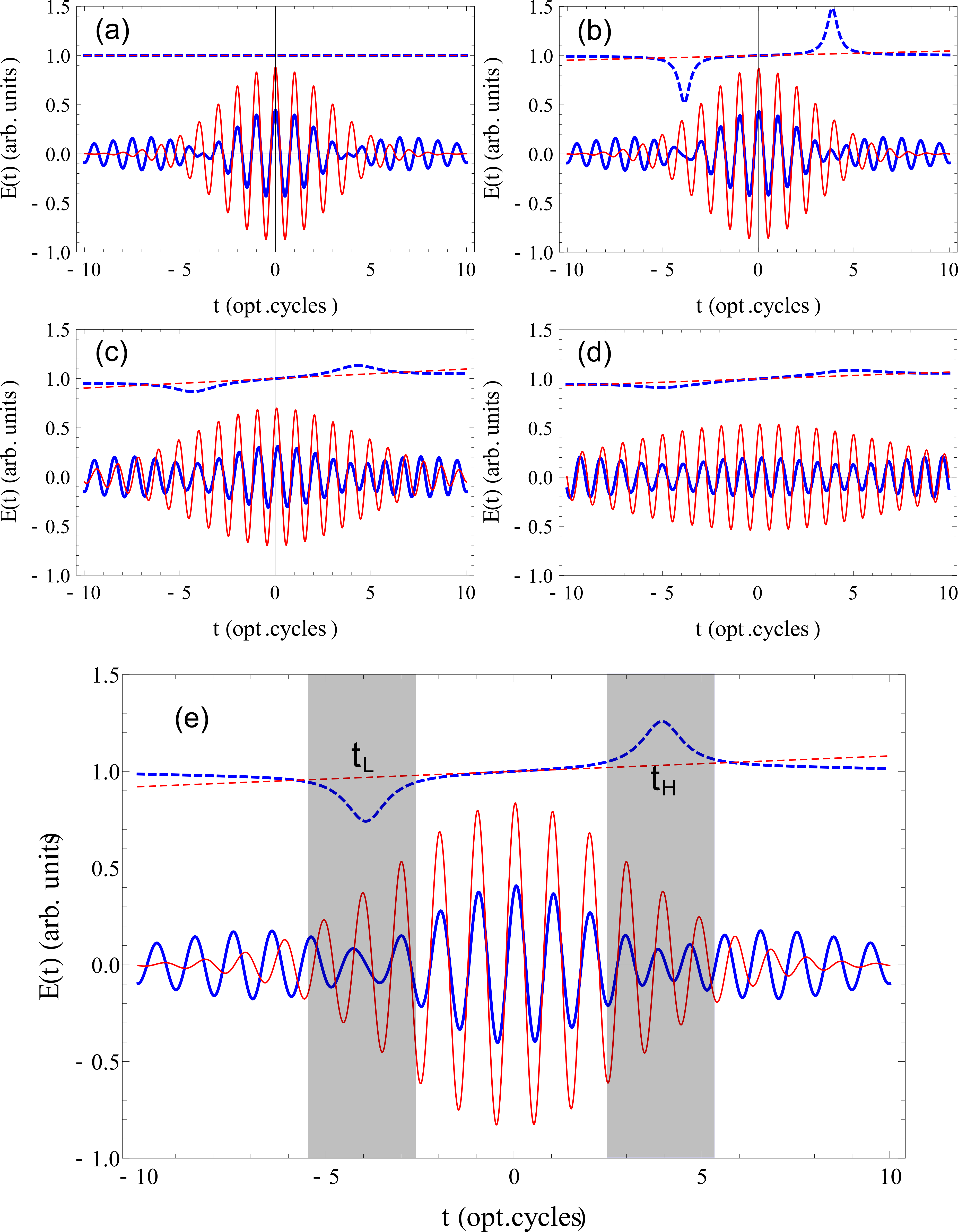}
\caption {
Instantaneous frequencies $\omega_{SB}^D(t)$ (blue dashed lines) and $\omega_{G}^D(t)$ (red dashed lines) normalized to the carrier frequency $\omega_0$. Panels (a), (b), (c), (d) and (d) correspond to different values of the parameter $\Omega$ that take in to account the propagation of the pulse through a dispersive medium: $\Omega=0,1,5,10$ and 2 opt. cycles$^2$, respectively. In addition, the corresponding fields $E_{SB}^D(t)$ (blue solid lines) and $E_{G}^D(t)$ (red solid lines) are shown.} \label{fig1}
\end{figure}

Our aim is to show the behavior of the instantaneous frequency of the SB pulse in the dispersive medium ($\omega_{SB}^D(t)$), and compare it with the behavior of that of the Gaussian pulse ($\omega_{G}^D(t)$). To this purpose, we calculated the expression in Eq.~\eqref{e6} for the set of values 
$\omega_0=2\pi$ and $\Delta\omega=0.5$ (both in units of $\frac{\mathrm{rad}}{\mathrm{opt. cycles}}$), which correspond to a Gaussian pulse $\tilde{E}_{G}(\omega)$ with a temporal FWHM, $\tau=\frac{2\sqrt{2ln2}}{\Delta\omega}\approx 5$ opt. cycles. These results are shown in Fig.~\ref{fig1}, where we have adopted the values $\alpha=1$ and $\beta=0.5$ for the synthesis parameters to obtain the SB pulse $\tilde{E}_{SB}(\omega)$, while
$\Omega$ was varied to take into account the effect of different media (GVD) and the propagation distance ($z$): $\Omega=0$ (Fig.~\ref{fig1}(a)),  $\Omega=1$ (Fig.~\ref{fig1}(b)), $\Omega=5$ (Fig.~\ref{fig1}(c)) and $\Omega=10$ (Fig.~\ref{fig1}(d)), all in units of opt. cycles$^2$. 
From these figures it can be seen that, while $\omega_{G}^D(t)$ has a linear behavior, $\omega_{SB}^D(t)$ is almost constant except within two temporal windows that are localized, symmetrically, on either sides of $t=0$. These local oscillations of the electric field $\tilde{E}_{SB}^D(\omega)$, have a frequency value that is below and above the minimum and maximum values reached by $\omega_{G}^D(t)$, and we will refer to these frequencies as $\omega_{low}$ and $\omega_{high}$, respectively.
Such oscillations tend to disappear when $\Omega$ increases, and $\omega_{SB}^D(t)$ converges to $\omega_G^D(t)$. 

A more detailed visualization of the phenomenon is shown in Fig.~\ref{fig1}(e). 
This figure is obtained for $\Omega = 2$ opt. cycles$^2$ and corresponds to high and low oscillations with frequency values $\omega_{high}\sim 1.25\omega_0$ and $\omega_{low}\sim0.75\omega_0$, localized at $t=t_L=-3.95$ opt. cycles and $t=t_H=3.95$ opt. cycles, respectively.
The shadow regions indicate the temporal windows where this phenomenon occurs. In addition, it can be observed how the period of the field $E_{SB}^D(t)$ (blue line) varies in relation to the period of $E_G^D(t)$ (red line). 
Finally, to 
estimate the ``weight" of these oscillations, 
we calculate the amplitude ratios between the different fields: $|E_{SB}^D(t_H)|/|E_{SB}^D(0)| = 0.192$, $|E_{SB}^D(t_H)|/|E_{G}^D(0)| = 0.094$, $|E_{SB}^D(t_H)|/|E_G(0)| = 0.089$.  As reference value, we can consider the weight in the Fourier spectrum of the Gaussian pulse at the frequency value   $\omega = 1.25\omega_0$, which is given by  $\tilde{E}_G(1.25\omega_0)/\tilde{E}_G(0) = e^{-(\frac{1.25\times 2\pi-2\pi}{0.5})^2}=e^{-\pi^2}=5.17\times 10^{-5}$. 
 From all the proposed ways to quantify the weight of the frequency $\omega_{high}$, this value is, at least, three order of magnitude greater than the value given by the Fourier spectrum. It should be note that $|E_{SB}^D(t_L)|=|E_{SB}^D(t_H)|$, so the analysis is also valid for the frequency $\omega_{low}$.   
%
\begin{figure}[h!]
\includegraphics[width=0.48\textwidth]{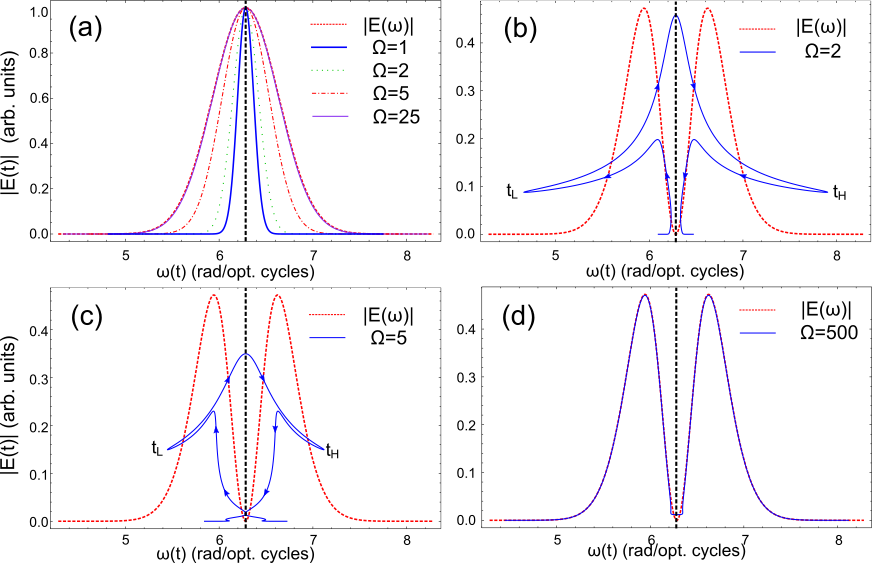}
\caption{
Parametric curves ($\omega(t),|E(t)|$) for a pulse propagating in a dispersive medium, and superimposed, the spectral amplitude $|\tilde{E}(\omega)|$ of the corresponding field. (a) Gaussian pulse $E_G^D(t)$. In panels (b), (c) and (d) the field considered is that of the SB pulse, $E_{SB}^D(t)$. The arrows in (b) and (c) indicate the evolution of
the parametric curve with time.} \label{fig2}
\end{figure}

In Fig.~\ref{fig2}, we show a set of parametric curves in the form ($\omega(t),|E(t)|$), which allow a general quantification of the oscillations represented by $\omega(t)$. In addition, the corresponding spectral amplitude $|\tilde{E}(\omega)|$ is superimposed in dashed red line. Figure~\ref{fig2}(a) corresponds to the parametric curves for the Gaussian pulse $E_G^D(t)$, propagating in a dispersive medium with $\Omega=1$ (blue line), $\Omega=2$ (green line), $\Omega=5$ (red line) and $\Omega=20$ (violet line) opt. cycles$^2$. From this figure, it can be seen how as the pulse propagates in the medium, the parametric curves converges to the spectral amplitude $|\tilde{E}_G(\omega)|$ (red dashed line). This result can be seen equivalent to obtain the temporal Fourier transform by the GVD of the medium \cite{Jannson1983}. 
In Fig.~\ref{fig2}(b), it is shown the parametric curve (blue line) for the SB pulse $E_{SB}^D(t)$, propagating in a dispersive medium with $\Omega=2$ opt. cycles$^2$ (same example of Fig. 1(e)), where the arrows indicate the evolution of the curve with time. 
From this figure, it can be seen how the curve cross the spectral amplitude $|\tilde{E}_{SB}(\omega)|$, and the maximum distance between both occurs for $t=t_L$ and $t=t_H$, denoting the appearance of two localized oscillations, of significant weight, that are clearly outside the Fourier spectrum of the pulse. From Fig.~\ref{fig2}(c), we can see a similar behavior, now for the case with $\Omega=5$ (same example of Fig.~\ref{fig1}(c)), being the maximum distance to $|\tilde{E}_{SB}(\omega)|$ shorter than in Fig.~\ref{fig2}(b).
Finally, in Fig. 2(d) it is shown a case with a great dispersion, $\Omega=500$ opt. cycles$^2$. Here, the shape of the parametric curve is quite similar to the shape of $|\tilde{E}_{SB}(\omega)|$. As in the case of Fig. 2(a), when the pulse propagates sufficiently in the medium, 
the temporal Fourier transform is obtained.   

An estimate of the ranges in which a behavior like that of Fig.~\ref{fig1}(e) is presents in a realistic situation, can be obtain, for example, by considered a fused silica glass whose GVD, at $\lambda_0=800$ nm ($T_0\approx 2.7$~ {\textit{fs}}), is 36.163~{\textit{fs}}$^2$/mm.
Since the dispersion length is defined as $L_D=\tau^2/\mathrm{GVD}$, then $\Omega=\frac{1}{2}\mathrm{GVD}\times z=\frac{1}{2}\tau^2\frac{z}{L_D}$.
In our case,
with $\Omega=2$ opt. cycles$^2$ and $\tau=\frac{2\sqrt{2ln2}}{\Delta\omega}\approx 5$ opt. cycles, we found that $z/L_D=2\Omega/\tau^2=4/25=0.16$, i.e., the pulse propagates in the medium, $z=0.16L_D\approx0.8$ mm. 

%
\begin{figure}[h!]
\includegraphics[width=0.4\textwidth]{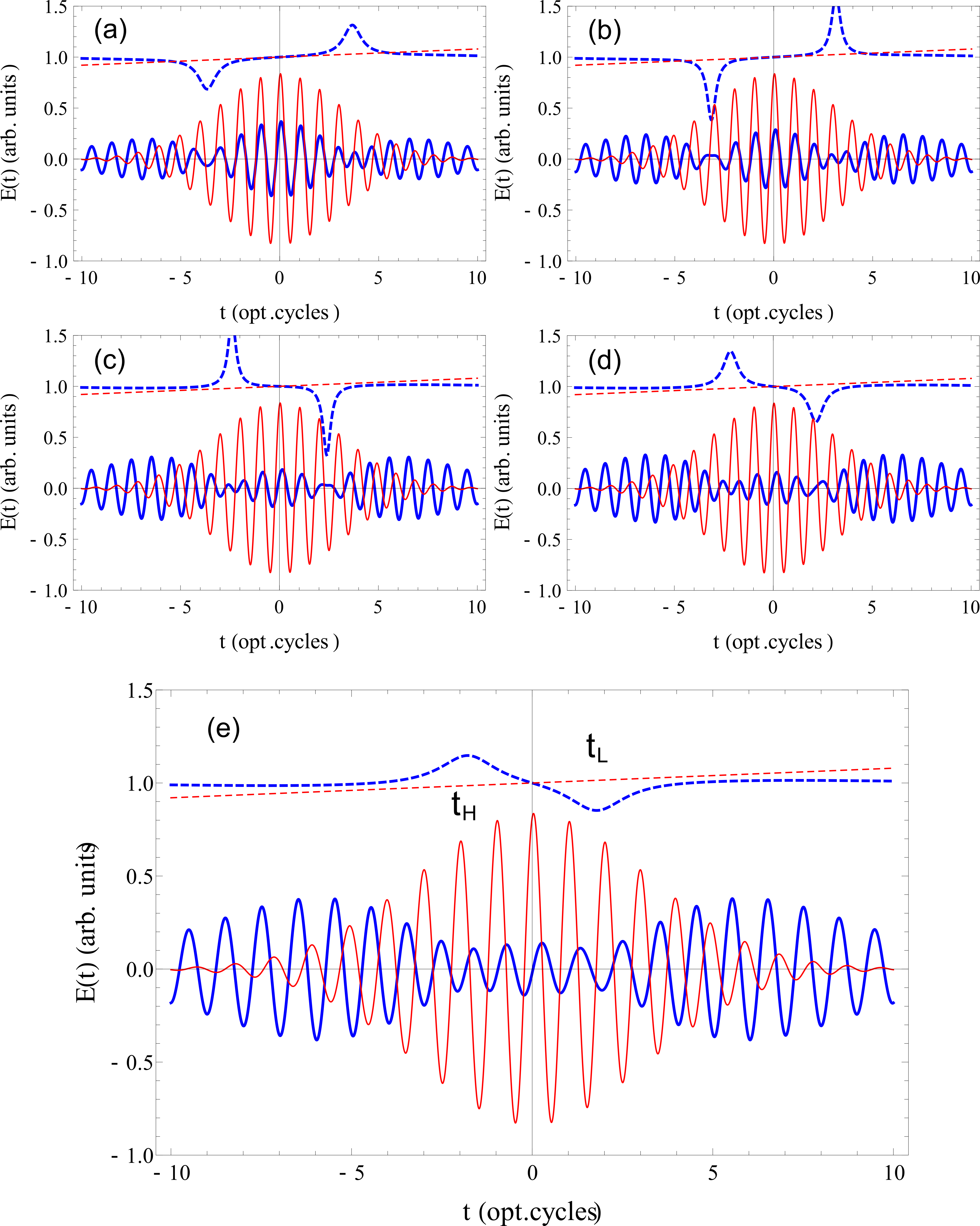}
\caption{Idem to Fig.\ref{fig1} but for different values of the synthesis parameter $\alpha$ of the SB pulse (Eq.~\eqref{e1}). Panels (a), (b), (c), (d) and (e) correspond to $\alpha$ = 1.1, 1.3, 1,6, 1,7 and 1,9, respectively. In all cases, the pulse propagates through a dispersive medium with $\Omega= 2$ opt. cycles$^2$.} \label{fig3}
\end{figure}

To study the effect of the dispersive medium on different SB pulses, in Fig.~\ref{fig3} we show the evolution of a pulse with the same synthesis parameter as the SB pulse analyzed up to this point, except for the value of the parameter $\alpha$. As we have shown in Ref.~\cite{Neyra22}, the temporal FWHM of the central lobe of the pulse given by Eq.~\eqref{e1}, decreases when the product $\alpha\times\beta$ increases, and converges to zero when $\alpha\times\beta\rightarrow 1$. In our case, since $\beta=0.5$, we have varied $\alpha$ in the range $1\leq\alpha< 1.9$. On the other hand, we consider that SB pulse propagates in a medium with $\Omega=2$opt. cycles$^2$. 
From Figs.~\ref{fig3}(a) ($\alpha=1.1$), ~\ref{fig3}(b) ($\alpha=1.3$), ~\ref{fig3}(c) ($\alpha=1.6$) and \ref{fig3}(d) ($\alpha=1.7$), it can be seen that the frequencies $\omega_{high}$ and $\omega_{low}$ (at $t_H$ and $t_L$, respectively) increase as $\alpha$ increases, up to a certain value for which the times $t_H$ and $t_L$ are reversed, and the opposite behavior is observed, i. e., the values of those frequencies begin to decrease when $\alpha$ increases. In Fig. 3 (e), we show the case with $\alpha=1.9$. 
It is observed a clear change in the behavior of $\omega_{SB}^D(t)$ with respect to lower values of $\alpha$. 
In a temporal region around $t=0$, $\omega_{SB}^D(t)$ can be approximate by a straight line with a negative slope, opposite to that of $\omega_G^D(t)$. 
The temporal position of the frequencies $\omega_{high}$ and $\omega_{low}$ are now inverted, 
i. e., $t_H < 0$ and $t_L >0$. 
In other temporal regions the value of $\omega_{SB}^D(t)$ remains constant. It is worth mention the similitude between the shape of $\omega_{SB}^D(t)$ and the shape of the instantaneous frequency given by the phenomenon known as {\it{self-phase modulation}}~\cite{neyra2016extending}. However, at difference of the phenomenon that we have shown here where the pulse propagates through a linear medium, the self-phase modulation phenomenon is a consequence of a non-linear response of the medium, where the phase acquired by the propagating pulse is dependent of its intensity.

\begin{figure}[h!]
\includegraphics[width=0.4\textwidth]{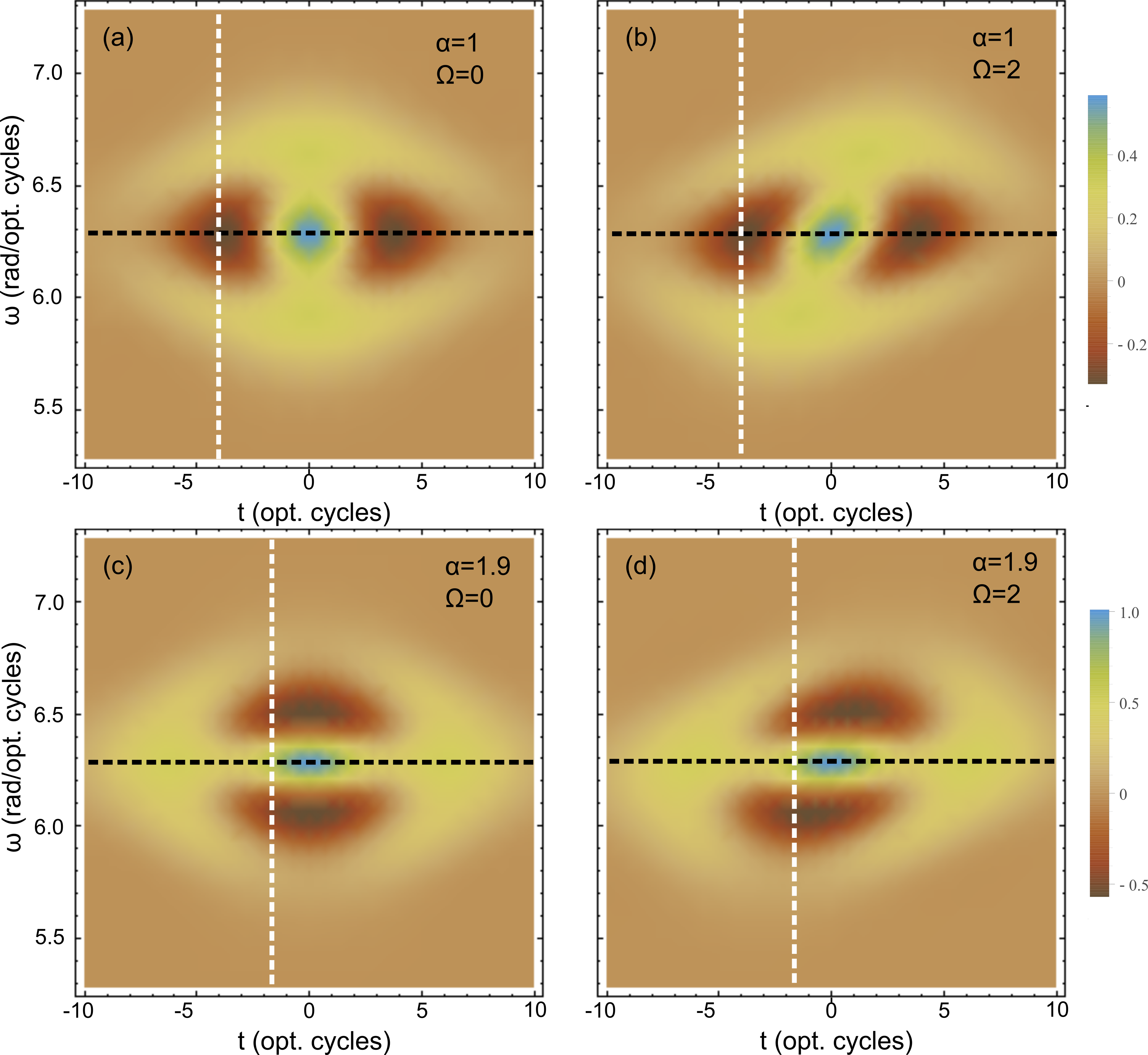}\\
\vspace{0.2cm}
%
\includegraphics[width=0.35\textwidth]{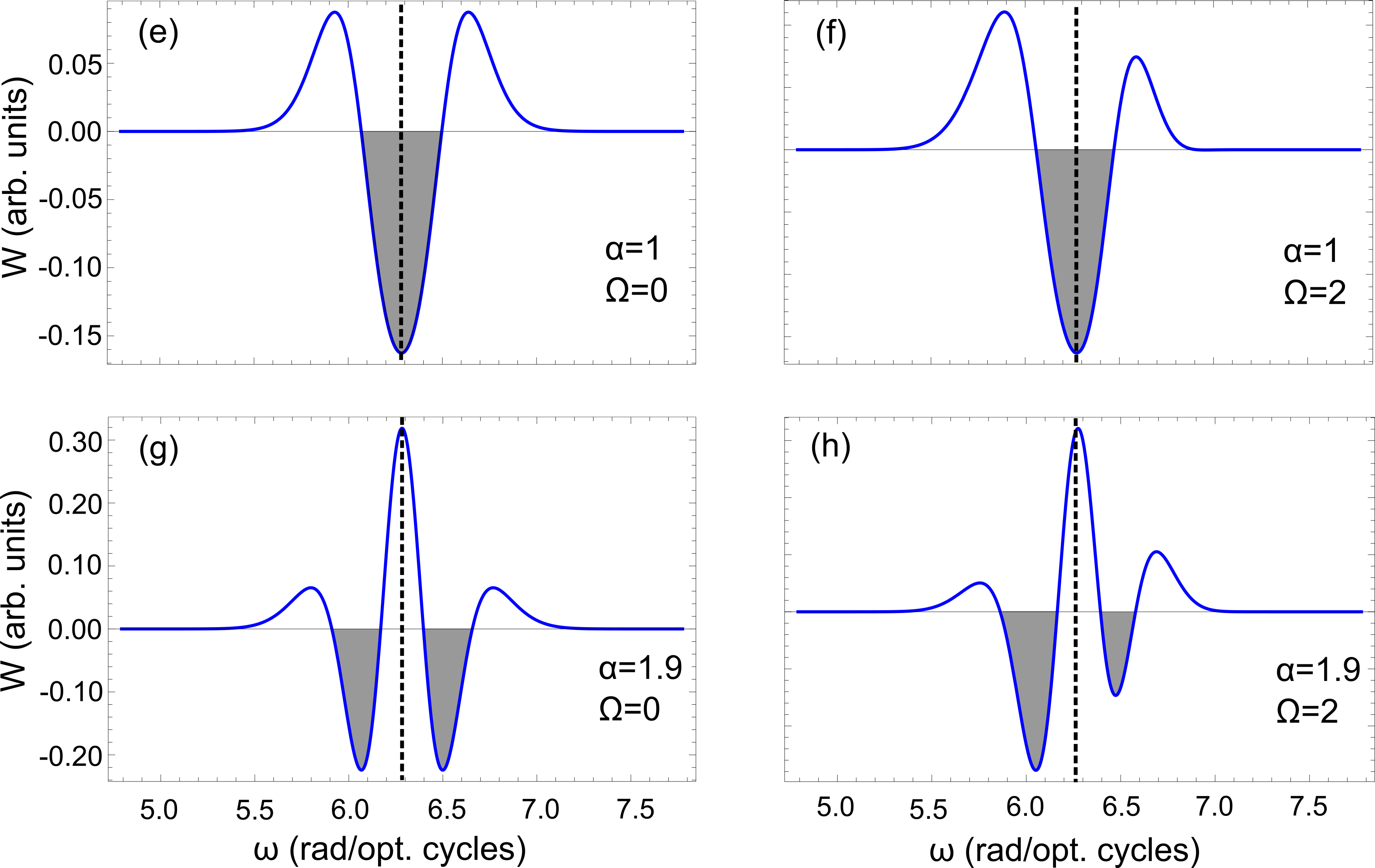}
\caption{Panels (a) to (d): Wigner function $W(t,\omega)$ of a propagating SB pulse. Panels (e) to (f): Cross section of the 2D map of $W(t,\omega)$ corresponding to the dashed white lines in panels (a) to (d), respectively.} \label{fig4}
\end{figure}

A different perspective of the underlying physics behind the phenomenon can be obtained from the analysis of the Wigner function distribution, which
allows us to decompose the \textit{superbandwidth} phenomenon in a time-frequency representation. To this end, in Fig.~\ref{fig4}, we show the Wigner distribution $W(t,\omega)$ for a SB pulse, as given by Eq.~\eqref{e5}. 
Two SB pulses, with a different value of the parameter $\alpha$,
are considered. When the SB pulses travel in a medium with $\Omega=0$, i.e., without dispersion, $W(t,\omega)$ exhibits a positive central lobe along two negative regions, horizontally or vertically aligned for the case with $\alpha=1$ (Fig.~\ref{fig4}(a)) or $\alpha=1.9$ (Fig.~\ref{fig4}(c)), respectively. %
%
In these cases, the negative values of $W(t,\omega)$ account for the destructive interference resulting from the coherent superposition of two Gaussian pulses~\cite{alonso2011wigner}, which results in the synthesis given by Eq.~\eqref{e2}. 
In addition, the Wigner distribution 
is symmetric with respect to both axes.
On the other hand, when the pulse propagates in a dispersive medium ($\Omega=2$ opt. cycles$^2$ for the examples that are shown in Figs.~\ref{fig4}(b) and \ref{fig4}(d) corresponding to $\alpha=1$ and $\alpha=1.9$, respectively), such symmetry is broken. 
This breaking of the mirror symmetry of $W(t,\omega)$ originates the time dependence of the instantaneous frequency $\omega(t)$. In Figs.~\ref{fig4}(e) to~\ref{fig4}(h), we show a cross section of the 2D maps in Figs.~\ref{fig4}(a) to~\ref{fig4}(d), respectively, for a given value of $t=t^*$ (dashed white lines). From these figures, it becomes clear that since $\omega(t^*)$ is the average frequency weighted by $W(t^*,\omega)$, when $\Omega=0$ the instantaneous frequency coincides with the carrier frequency ($\omega(t^*) = \omega_0$), independently of the negative values of $W(t^*,\omega)$ (Figs.~\ref{fig4}(e) and ~\ref{fig4}(g)). For $\Omega\neq 0$, the asymmetry of $W(t^*,\omega)$ causes the time-dependence in $\omega(t)$, which can now reach 
higher and lower values than those given by the Fourier spectrum, as a consequence of the negative values of $W(t^*,\omega)$~\cite{berry2014superoscillations} (otherwise, a positive function distribution always has a first moment whose value is between the minimum and maximum values of the variable).
%
%
%

\begin{figure}[h!]
\includegraphics[width=0.48\textwidth]{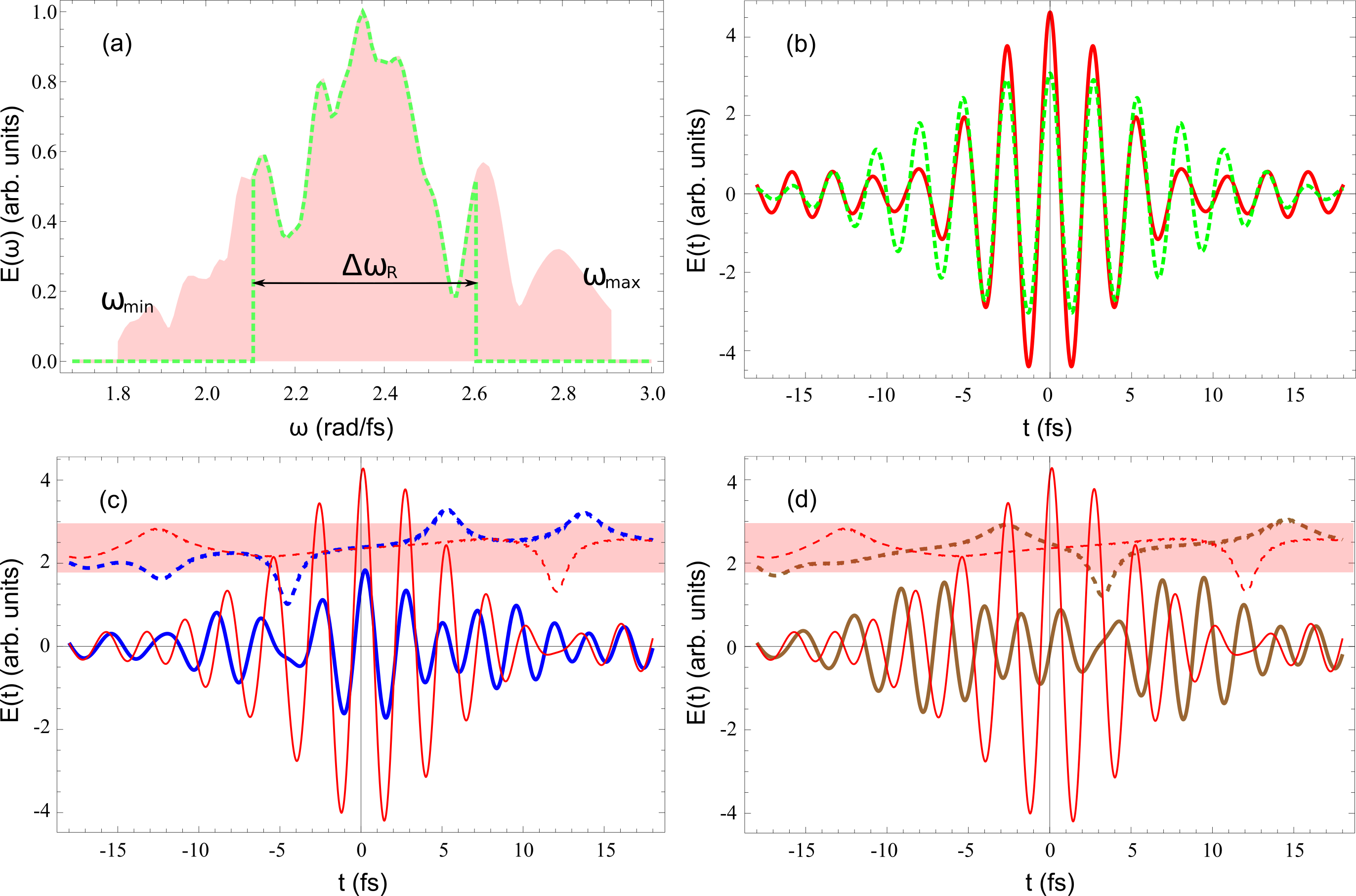}
\caption{(a) Fourier spectrum of the non-Gaussian pulse $\tilde{E}_{NG} (\omega)$ (pink region), and response of the rectangular spectral filter to obtain the pulse $\tilde{E}_{R}(\omega)$ (green dashed lines). (b) Pulses $E_{NG}(t)$ (red solid line) and $E_{R}(t)$ (green dashed line). (c) Instantaneous frequencies $\omega_{SB}^D(t)$ (blue dashed line) and $\omega_{NG}^D(t)$ (red dashed line) for a synthesis parameter $\alpha=0.85$, when the propagation length in the dispersive medium is $z=300\mu $m. The pink region indicates the spectral content of $E_{NG}(t)$ (see Fig.~\ref{fig5}(a)). In addition, the propagating pulses $E_{SB}^D(t)$ (blue solid line) and  $E_{NG}^D(t)$ (red solid line) are shown. (d) Idem (c) but for $\alpha=1.37$. Here, $\omega_{SB}^D(t)$ is display in brown dashed line and $E_{SB}^D(t)$ in brown solid line.} \label{fig5}
\end{figure}

To conclude, we highlight that the phenomena presented in this work  
extend to SB pulses synthesized from the coherent superposition of non-Gaussian ultra-short pulses, with a complex frequency spectrum.
As a concrete example, in Fig.~\ref{fig5}(a), we show the Fourier spectrum of a non-Gaussian pulse $\tilde{E}_{NG}(\omega)$ 
(pink region), which is centered at $\omega_C\approx2.35$ rad/\textit{fs} ($\lambda_C\approx 800$ nm), and is limited between $\omega_{min}\approx1.8$ and $\omega_{max}\approx2.9$ rad/\textit{fs}. The dashed green lines indicate the bandwidth $\Delta\omega_{R}$ of the spectral rectangular filter considered to obtain a second non-Gaussian pulse $\tilde{E}_{R}(\omega)$, 
which subsequently interferes with $\tilde{E}_{NG}(\omega)$.
Figure~\ref{fig5}(b) corresponds to the temporal representation of the original pulse $E_{NG}(t)$ (red solid line), and the pulse $E_{R}(t)$ (green dashed line) broadened as a consequence of the spectral filtering. As previously, in the frequency domain the SB pulse is given by $\tilde{E}_{SB}(\omega) = \tilde{E}_{NG}(\omega)-\alpha\tilde{E}_{R}(\omega)$ (see Eq.~\eqref{e1}), and different synthesis result depending on the amplitude ratio $\alpha$ between both fields.
%
%
Without loss of generality, we have considered that the SB pulse propagates through a fused silica glass, so that $E_{SB}^D(t) = \mathcal{F}[\tilde{E}_{SB}(\omega)e^{-iK(\omega)\times z}]$, with $K(\omega)=\frac{\omega }{c}n_{glass}(\omega)$, for which a fifth order fit function in $\omega$ was considered, being $n_{glass}(\omega)$ the refractive index of the medium~\cite{diels2006ultrashort}. 
%
In Fig.~\ref{fig5}(c), corresponding to $\alpha=0.85$ and $z=300\mu$m, the instantaneous frequency $\omega_{SB}^D(t)$ reveals the presence of two local oscillations beyond the Fourier spectrum of the pulse $E_{NG}(t)$, in the central region where the frequency $\omega_{NG}^D(t)$ remains almost linear.
Similar to what was seen in Fig.~\ref{fig1},
$\omega_{SB}^D(t)$ has a behavior consistent with a positive dispersive medium, $\mathrm{GVD} > 0$. In the present example, unlike the former one, a second pair of oscillations appear on either side of the central region, where $\omega_{NG}^D(t)$ deviates from the linear behavior and also shows two side oscillations. However, these secondary oscillations have a negligible weight compared to the primary ones.
Finally, in Fig.~\ref{fig5}(d) which corresponds to $\alpha=1.37$ and $z=300\mu$m,
the primary local oscillations exhibit the changing behavior 
previously observed in Fig.~\ref{fig3}, that is, they travel as in a medium with $\mathrm{GVD} < 0$. 
%

{\textit{Outlook}}.---We have studied the superbandwidth phenomenon in laser pulses through the propagation characteristics of SB pulses in a dispersive medium. The appearance of local oscillations of the electric field, characterized by the instantaneous frequency $\omega(t)$, 
strengthens the concept of superbandwidth indicating that the SB pulse interacts, locally, as if it had frequencies higher and lower than those present in its Fourier spectrum. For some values of the pulse synthesis parameters, such local oscillations
behave as if they travel in a medium with positive or negative GVD, 
beyond the dispersion properties of the material. At the best of our knowledge, this is the first time that a phenomenon with this characteristics is observed  in a linear medium.   

Besides the direct application that these results could have for coherent control~\cite{rybka2016sub,he2019coherently,koong2021coherent}, pulse shaping~\cite{weiner2011ultrafast,manzoni2015coherent,ridente2022electro}, signal processing~\cite{krausz2014attosecond,lee2014superoscillations,lee2014superoscillations1} and ultrafast spectroscopy ~\cite{maiuri2019ultrafast,cavalieri2007attosecond,shah2013ultrafast}, the concept of superbandwidth of a band-limited function can be extended to other waves phenomena. Moreover, the spatio-temporal coupling in ultrashort laser pulses could be the key to isolate, spatially, the local oscillations of the electric field given by its superbandwidth~\cite{vincenti2012attosecond,akturk2010spatio}.   

Since this paper focused on the most relevant results of our work, a complete analysis of the phenomenon will be addressed in a future work, where it will be discussed how $\omega_{SB}^D(t)$ and the temporal FWHM of the central lobe depend on the pulse synthesis parameters 
and the propagation length. 
Furthermore, it is of interest to carry out a deeper analysis of the physics behind the change in the regimen where $\omega_{SB}^D(t)$ behave as if have positive dispersion or negative dispersion.




\bibliography{main}

\end{document}